\documentclass[twocolumn,pra,showpacs,showkeys,preprintnumbers,amsmath,amssymb,floatfix]{revtex4}
\usepackage{graphicx}
\usepackage{dcolumn}% Align table columns on decimal point
\usepackage{bm,bbm,mathrsfs}% bold math
\renewcommand{\theequation}{\arabic{section}.\arabic{equation}}

\newcommand{\ot}[0]{\otimes}
\newcommand{\avg}[3]{\big<#1\big|#2\big|#3\big>}

\newcommand{\wektor}[1]{\boldsymbol{\mathrm{#1}}}
\newcommand{\wersor}[1]{\boldsymbol{\mathrm{\hat{#1}}}}

\newcommand{\braket}[2]{\big<#1\big|#2\big>}
\newcommand{\calkaob}[2]{\int_{\mathbb{R}^{3}}d^{3}
\wektor{#1}#2}
\newcommand{\calkapow}[2]{\oint_{\mathrm{4\pi}}d^{2}
{\hat{\wektor{#1}}}#2}
\def\LRA{\mathop{-\!\!\!-\!\!\!
\longrightarrow}\nolimits}
\newcommand{\Amplituda}[0]{A_{fi}}
\newcommand{\Amplitudaa}[0]{\mathcal{A}_{fi}}
\newcommand{\Amplitudaaa}[0]{\mathscr{A}_{fi}}

\newcommand{\mac}[1]{\mathsf{#1}}
\newcommand{\jed}[0]{\mathbbm{1}}
\begin{document}
\title{Scattering of Dirac particles from non-local separable
potentials:\\ The eigenchannel approach

}% Force line breaks with \\

\author{Remigiusz Augusiak}
\email{remik@mif.pg.gda.pl} \affiliation{Department of Theoretical
Physics and Quantum Informatics,\\ Faculty of Applied Physics and
Mathematics, Gda\'nsk University of Technology,\\ Narutowicza
11/12, PL 80-952 Gda\'nsk, Poland}

\date{\today}

\begin{abstract}
An application of the new formulation of the eigenchannel method
[R. Szmytkowski, Ann. Phys. (N.Y.) {\bf 311}, 503 (2004)] to
quantum scattering of Dirac particles from non-local separable
potentials is presented. Eigenchannel vectors, related directly to
eigenchannels, are defined as eigenvectors of a certain weighted
eigenvalue problem. Moreover, negative cotangents of
eigenphase-shifts are introduced as eigenvalues of that spectral
problem. Eigenchannel spinor as well as bispinor harmonics are
expressed throughout the eigenchannel vectors. Finally, the
expressions for the bispinor as well as matrix scattering
amplitudes and total cross section are derived in terms of
eigenchannels and eigenphase-shifts. An illustrative example is
also provided.
\end{abstract}

%\keywords{Eigenchannels, potential scattering,
%separable potentials.}%Use showkeys class option if keyword
                              %display desired
\maketitle

\section{Introduction}

Recently, Szmytkowski \cite{Szmytkowski} proposed a new formulation
of the eigenchannel method for quantum scattering from Hermitian
short-range potentials, different from that presented by Danos and
Greiner \cite{Danos}. Some ideas leading to this method were drawn
from works on electromagnetism theory by Garbacz \cite{Garbacz1} and
Harrington and Mautz \cite{Harrington1}. This method was further
extended to the case of zero-range potentials for Schr\"odinger
particles by Szmytkowski and Gruchowski \cite{Szmytkowski2} and then
for Dirac particles by Szmytkowski \cite{Szmytkowski3} (see also
\cite{Szmytkowski5}).

On the other hand, it is the well-known fact that separable
potentials, since they provide analytical solutions to the
Lippmann-Schwinger equations \cite{LippSchw}, have found
applications in many branches of physics, both in the
non-relativistic and relativistic cases \cite{Zast}. (It should be
noted that much larger effort has been devoted to the separable
potentials in the non-relativistic regime.) Especially, their
utility was confirmed in nuclear physics by successful use for
describing nucleon-nucleon interactions \cite{NN}. Moreover,
methods allowing one to approximate an arbitrary non-local
potential by a separable one are known \cite{metody}.

In view of what has been said above, it seems interesting to pose
the question: {\it how does the new method apply to quantum
scattering from non-local separable potentials?} Partially, the
answer was given by the author by applying the method to quantum
scattering of Schr\"odinger particles from separable potentials
\cite{moja}. In the present contribution, we extend considerations
from \cite{moja} to the case of Dirac particles.

This paper is organized as follows. In Section 2 some facts and
notions from the theory of potential scattering of Dirac particles
(see \cite{Thaller}) are provided.
In Section 3 we concentrate on the special class of non-local
potentials, namely, separable potentials. In this context,
expressions for the bispinor as well as matrix scattering
amplitudes are provided. Section 4 contains main ideas and
results. Here, we define {\it eigenchannel vectors}, directly
related to eigenchannels, as solutions to a certain weighted
eigenproblem. Moreover, we introduce eigenphase-shifts, relating
them to eigenvalues of this spectral problem.  Within this
approach, we also calculate expressions for the scattering
amplitude and the average total cross section. In Section 5,
scattering from a rank one delta-like separable potential is
discussed as an illustrative example. The paper ends with two
appendices.

\section{Quantum scattering of Dirac particles from non-local potentials}
\label{SecII}

Let us assume that a free Dirac particle of energy $E$ (with
$|E|>mc^{2}$) described by the following monochromatic plane wave
\begin{equation}\label{I.1}
\phi_{i}(\wektor{r})\equiv\braket{\wektor{r}}{\wektor{k}_{i}\chi_{i}}=
U_{i}(\wektor{k}_{i})e^{i\wektor{k}_{i}\cdot\wektor{r}},
\end{equation}
where
\begin{equation}\label{I.2}
U_{i}(\wektor{k}_{i})=\frac{1}{\sqrt{1+\varepsilon^{2}}} \left(
\begin{array}{c}
\chi_{i} \\*[0.2ex]
\displaystyle\varepsilon\wektor{\sigma}\cdot\wersor{k}_{i}\,\chi_{i}
\end{array}
\right),
\end{equation}
\begin{equation}
%\epsilon=\frac{\varepsilon}{k}, \quad
\varepsilon=\sqrt{\frac{E-mc^{2}}{E+mc^{2}}}
\end{equation}
is being scattered from a non-local potential given by a kernel
$\mathsf{V}(\wektor{r},\wektor{r}')$, which in general may be a
$4\times 4$ matrix. In the above equation, $\chi_{i}$ stands for a
normalized pure spin-$\frac{1}{2}$ state belonging to
$\mathbb{C}^{2}$. Orientation of the spin in $\mathbb{R}^{3}$ will
be denoted by $\wektor{\nu}_{i}$ and is related to $\chi_{i}$ by
$\wektor{\nu}_{i}=\chi^{\dagger}_{i}\wektor{\sigma}\chi_{i}$,
where $\wektor{\sigma}$ is a vector consisting of the standard
Pauli matrices, i.e.,
\begin{equation}\label{I.3}
\wektor{\sigma}= \left[ \left(
\begin{array}{cc}
0 & 1\\
1 & 0
\end{array}
\right), \left(
\begin{array}{cc}
0 & -i\\
i & 0
\end{array}
\right), \left(
\begin{array}{cc}
1 & 0\\
0 & -1
\end{array}
\right) \right].
\end{equation}
Moreover, $\wektor{p}_{i}=\hbar\wektor{k}_{i}$ is a momentum of
the incident particle and $k$ denotes the Dirac wave number and is
given by
\begin{equation}\label{DiracWaveNumber}
k=\mathrm{sgn}(E)\sqrt{\frac{E^{2}-\left(mc^{2}\right)^{2}}{c^{2}\hbar^{2}}}.
\end{equation}
Thereafter, we shall consider only Hermitian potentials, i.e.,
those with kernels obeying
$\mathsf{V}(\wektor{r},\wektor{r}')=\mathsf{V}^{\dagger}(\wektor{r}',\wektor{r})$.

For this scattering process we may write the Lippmann-Schwinger
equation \cite{LippSchw} of the form
\begin{eqnarray}\label{I.4}
&&\hspace{-1.4cm}\psi(\wektor{r})=\phi_{i}(\wektor{r})\nonumber\\
&&\hspace{-1cm}-\calkaob{r}'\calkaob{r}''\:G(E,\wektor{r},\wektor{r}')\mathsf{V}(\wektor{r}',\wektor{r}'')
\psi(\wektor{r}'').
\end{eqnarray}
Function $G(E,\wektor{r},\wektor{r}')$ appearing above is the
relativistic free--particle outgoing Green function given by
\begin{equation}\label{I.6}
G(E,\wektor{r},\wektor{r}')= \frac{1}{4\pi c^{2}\hbar^{2}}\left(
-ic\hbar\wektor{\alpha}\cdot\wektor{\nabla}+\beta mc^{2}+E\jed_{4}
\right)
\frac{e^{ik|\wektor{r}-\wektor{r}'|}}{|\wektor{r}-\wektor{r}'|},
\end{equation}
and formally is a kernel of the relativistic outgoing Green
operator defined as
\begin{equation}\label{I.7}
\hat{G}(E)=\lim_{\epsilon\downarrow
0}[\hat{\mathcal{H}}_{0}-E-i\epsilon]^{-1},
\end{equation}
with
$\hat{\mathcal{H}}_{0}=-ic\hbar\wektor{\alpha}\cdot\wektor{\nabla}+
\beta mc^{2}$ being a Dirac free--particle Hamiltonian. Here
\begin{equation}\label{I.5}
\wektor{\alpha}= \left(
\begin{array}{cc}
0 & \wektor{\sigma}\\
\wektor{\sigma} & 0
\end{array}
\right), \qquad \beta= \left(
\begin{array}{cc}
\jed_{2} & 0\\
0 & -\jed_{2}
\end{array}
\right),\qquad \jed_{2}=\left(
\begin{array}{cc}
1 & 0\\
0 & 1
\end{array}
\right)
\end{equation}
and $\jed_{4}=\jed_{2}\ot\jed_{2}$.
%and $\sf{0}_{2}$ is a $2\times2$ zero matrix.
It is worth noticing that within the relativistic
regime the Green function (\ref{I.6}) is a $4\times 4$ matrix.

For purposes of further analysis, it is useful to introduce the
following projector:
\begin{equation}\label{I.8}
\mathcal{P}(\wektor{k})=\frac{c\hbar
\wektor{\alpha}\cdot\wektor{k}+\beta mc^{2}+E\jed_{4}}{2E},
\end{equation}
which, as one can immediately infer, may be decomposed in the
following way
\begin{eqnarray}\label{I.9}
\mathcal{P}(\wektor{k})&=&\Theta_{+}(\wektor{k})
\Theta_{+}^{\dagger}(\wektor{k})+\Theta_{-}(\wektor{k})\Theta_{-}^{\dagger}(\wektor{k})\nonumber\\
&=&\frac{1}{1+\varepsilon^{2}} \left(
\begin{array}{cc}
\mathbbm{1}_{2} & \varepsilon \wektor{\sigma}\cdot\wersor{k}\\
\varepsilon \wektor{\sigma}\cdot\wersor{k} &
\varepsilon^{2}\mathbbm{1}_{2},
\end{array}
\right)
\end{eqnarray}
with $\Theta_{\pm}(\wektor{k})$ being defined as
\begin{equation}\label{I.10}
\Theta_{\pm}(\wektor{k})= \frac{1}{\sqrt{1+\varepsilon^{2}}}\left(
\begin{array}{c}
\theta_{\pm}\\
\varepsilon\wektor{\sigma}\cdot\wersor{k}\,\theta_{\pm}
\end{array}
\right).
\end{equation}
Spinors $\theta_{\pm}$ constitute an arbitrary orthonormal basis
in $\mathbb{C}^{2}$, i.e.,
$\theta_{s}^{\dagger}\theta_{t}=\delta_{st}$ $(s,t=-,+)$ and
$\sum_{s=-}^{+}\theta_{s}\theta_{s}^{\dagger}=\jed_{2}$. What is
important for further considerations, the matrix (\ref{I.8})
possesses the obvious property that
$\mathcal{P}(\wektor{k})\Theta_{\pm}(\wektor{k})=\Theta_{\pm}(\wektor{k})$
and therefore
\begin{equation}\label{I.11}
\mathcal{P}(\wektor{k}_{i})U_{i}(\wektor{k}_{i})=U_{i}(\wektor{k}_{i}).
\end{equation}
We shall be exploiting this property in later analysis.

Considering scattering processes we usually tend to find
expressions for a scattering amplitude and various cross sections.
To this aim we need to find an asymptotic behavior of the
relativistic outgoing Green function. From Eq. (\ref{I.6}), using
the projector (\ref{I.8}), we have
\begin{equation}\label{I.12}
G(E,\wektor{r},\wektor{r}')\stackrel{r\to\infty}{\sim}
\frac{E}{2\pi c^{2}\hbar^{2}}
\mathcal{P}(\wektor{k}_{f})\frac{e^{ikr}}{r}
e^{-i\wektor{k}_{f}\cdot\wektor{r}'},
\end{equation}
where $\wektor{k}_{f}=k\wektor{r}/r$ is a wave vector of the
scattered particle. Notice that due to the fact that we deal with
elastic processes $|\wektor{k}_{i}|=|\wektor{k}_{f}|=k$. After
application of Eq. (\ref{I.12}) to Eq. (\ref{I.4}), we obtain
\begin{equation}\label{I.13}
\psi(\wektor{r})\stackrel{r\to\infty}
{\sim}\underset{r\to\infty}{\mathrm{asymp}}\,  \phi_{i}(\wektor{r})
+\Amplituda\frac{e^{ikr}}{r},
\end{equation}
where $\Amplituda$ is the bispinor scattering amplitude and is
defined through the relation
\begin{eqnarray}\label{I.14}
&&\hspace{-1.5cm}\Amplituda=-\frac{E}{2\pi
c^{2}\hbar^{2}}\mathcal{P}(\wektor{k}_{f})\nonumber\\
&&\hspace{-1cm}\times\calkaob{r}'\calkaob{r}''\,e^{-i\wektor{k}_{f}\cdot\wektor{r}'}
\mathsf{V}(\wektor{r}',\wektor{r}'')\psi(\wektor{r}'')
\end{eqnarray}
and, in general, is of the form
\begin{equation}\label{I.15}
\Amplituda=\frac{1}{\sqrt{1+\varepsilon^{2}}}\left(
\begin{array}{c}
\chi_{f}\\
%(\wektor{k}_{i},\wektor{k}_{f})\\
\varepsilon\wektor{\sigma}\cdot\wersor{k}_{f}\chi_{f}
%(\wektor{k}_{i},\wektor{k}_{f})
\end{array}
\right).
\end{equation}
Here $\chi_{f}$ is a spinor transformed from the initial spinor
$\chi_{i}$ by the scattering process. Vector
$\wektor{\nu}_{f}=(\chi_{f}^{\dagger}\wektor{\sigma}\chi_{f})/(\chi_{f}^{\dagger}\chi_{f})$
responds for an orientation of the spin of the scattered particle.
Therefore let us assume that there exist a matrix, such that
$\chi_{f}=\Amplitudaaa\chi_{i}$. Then it is easy to verify that
the bispinor scattering amplitude may be written in the form
\begin{equation}\label{I.16}
\Amplituda=\Amplitudaa U_{i}(\wektor{k}_{i}),
\end{equation}
where the matrix $\Amplitudaa$ is related to $\Amplitudaaa$ by
\begin{equation}\label{I.17}
\Amplitudaa=\frac{1}{1+\varepsilon^{2}}  \left(
\begin{array}{cc}
\Amplitudaaa &\quad
\varepsilon\Amplitudaaa\wektor{\sigma}\cdot\wersor{k}_{i}\\
\varepsilon\wektor{\sigma}\cdot\wersor{k}_{f}\Amplitudaaa &\quad
\varepsilon^{2}\wektor{\sigma}\cdot\wersor{k}_{f}\Amplitudaaa\wektor{\sigma}\cdot\wersor{k}_{i}
\end{array}
\right).
\end{equation}
Henceforth matrices $\Amplitudaa$ and $\Amplitudaaa$ will be
called the matrix scattering amplitudes. The differential cross
section for scattering from the direction $\wektor{k}_{i}$ and the
spin arrangement $\wektor{\nu_{i}}$ onto $\wektor{k_{f}}$ and
$\wektor{\nu_{f}}$ is defined as
\begin{equation}\label{I.18}
\frac{\mathrm{d}\sigma}{\mathrm{d}\Omega_{f}}
%(\wektor{k}_{f}\leftarrow\wektor{k}_{i})
=\chi_{f}^{\dag}\chi_{f}=\chi_{i}^{\dagger}
\Amplitudaaa^{\dagger}\Amplitudaaa\chi_{i},
\end{equation}
Subsequently, after integration the above over all the directions
of $\wektor{k}_{f}$, we arrive at the total cross section
\begin{equation}\label{I.19}
\sigma(\wektor{k}_{i},\wektor{\nu}_{i})=
\calkapow{k}_{f}\:\chi_{f}^{\dag}\chi_{f}.
\end{equation}
Finally, averaging over all directions of incidence
$\hat{\wektor{k}}_{i}$ and the initial spin orientation
$\hat{\wektor{\nu}}_{i}$, one finds the average total cross
section
\begin{equation}\label{I.20}
\sigma_{t}(E)=
\frac{1}{(4\pi)^{2}}\calkapow{k}_{i}\calkapow{\nu}_{i}\calkapow{k}_{f}
\:\chi_{f}^{\dag}\chi_{f}.
\end{equation}
Obviously all the mentioned cross sections may be expressed in terms
of all the scattering amplitudes $\Amplituda$, $\Amplitudaa$, and
$\Amplitudaaa$.

%%%%%%%%%%%%%%%%%%%%%%%%%%%%%%%%%%%%%%%%%%%%%%%%%%%%%%%%%%%%%%%%%%%%%%%%
%%%%%%%%%%%%                                               %%%%%%%%%%%%%
%%%%%%%%%%%%   Section II                                  %%%%%%%%%%%%%
%%%%%%%%%%%%                                               %%%%%%%%%%%%%
%%%%%%%%%%%%%%%%%%%%%%%%%%%%%%%%%%%%%%%%%%%%%%%%%%%%%%%%%%%%%%%%%%%%%%%%

\section{Special class of non--local separable potentials}
\setcounter{equation}{0} \noindent\indent In this section we
employ the above considerations to the special class of non--local
separable potentials. As previously mentioned, such a class of
potentials allows to find solutions to the Lippmann--Schwinger
equations in an analytical way.

Consider the following class of potential kernels:
%
%POT SEP RELATYWISTYCZNIE
\begin{equation}\label{II.1}
\mathsf{V}(\wektor{r},\wektor{r}')=\sum_{\mu}\omega_{\mu}
\mathsf{u}_{\mu}(\wektor{r})\mathsf{u}_{\mu}^{\dag}(\wektor{r}')
%\qquad\left( \displaystyle\forall\mu:\;\;\omega_{\mu}\in\mathbb{R}\setminus\{0\} \right),
\end{equation}
where it is assumed that in general $\mu$ may denote the arbitrary
finite set of indices, i.e., $\mu=\{\mu_{1},\ldots,\mu_{k}\}$ and
all the coefficients $\omega_{\mu}$ different from zero. Functions
$\mathsf{u}_{\mu}(\wektor{r})$ are assumed to be four--element
columns.

Substitution of Eq. (\ref{II.1}) to Eq. (\ref{I.4}) leads us to
the Lippmann--Schwinger equation for the separable potentials:
\begin{eqnarray}\label{II.2}
&&\hspace{-0.2cm}\psi(\wektor{r})=\phi_{i}(\wektor{r})-\sum_{\mu}\omega_{\mu}
\calkaob{r}'\,G(E,\wektor{r},\wektor{r}')\mathsf{u}_{\mu}(\wektor{r}')
\nonumber\\
&&\hspace{1cm}\times\calkaob{r}''\,\mathsf{u}_{\mu}^{\dag}(\wektor{r}'')\psi(\wektor{r}''),
\end{eqnarray}
which may be equivalently rewritten as a set of linear algebraic
equations. Indeed, using the Dirac notation one finds
\begin{equation}\label{II.3}
\sum_{\mu} \left[ \delta_{\nu\mu}+
\avg{\mathsf{u}_{\nu}}{\hat{G}(E)}{\mathsf{u}_{\mu}} \omega_{\mu}
\right]
\braket{\mathsf{u}_{\mu}}{\psi}=\braket{\mathsf{u}_{\nu}}{\phi_{i}}.
\end{equation}
For further convenience we introduce the following notations
\begin{eqnarray}\label{II.4}
&&\big<\mathsf{\mathsf{u}}\big|\varphi\big>= \left(
\begin{array}{c}
\big<\mathsf{u}_{1}\big|\varphi\big>\\*[0.5ex]
\big<\mathsf{u}_{2}\big|\varphi\big>\\
\vdots
\end{array}
\right),\nonumber\\
&&\big<\varphi\big|\mathsf{\mathsf{u}}\big>=\big<\mathsf{u}\big|\varphi\big>^{\dagger}=
\left( \big<\varphi\big|\mathsf{u}_{1}\big>\;
\big<\varphi\big|\mathsf{u}_{2}\big>\; \dots \right).
\end{eqnarray}
Consequently, we may rewrite Eq. (\ref{II.3}) as a matrix equation
$(\mathbbm{1}+\mathsf{G}\mathsf{\Omega})
\big<\mathsf{u}\big|\psi\big>=\big<\mathsf{u}\big|\phi_{i}\big>$
or equivalently as
\begin{equation}\label{II.6}
\big<\mathsf{u}\big|\psi\big>=(\mathbbm{1}+\mathsf{G}\mathsf{\Omega})^{-1}
\big<\mathsf{u}\big|\phi_{i}\big>,
\end{equation}
with $\mathsf{G}$ being a matrix composed of the elements
$\avg{\mathsf{u}_{\nu}}{\hat{G}(E)}{\mathsf{u}_{\mu}}$ and
$\sf{\Omega}=\mathrm{diag}[\omega_{\mu}]$. Similarly, substituting
Eq. (\ref{II.1}) to Eq. (\ref{I.14}) and again using Eq.
(\ref{I.8}), we arrive at the bispinor scattering amplitude for the
separable potentials in the form
\begin{eqnarray}\label{II.7}
&&\hspace{-1.5cm}\Amplituda=\frac{-E}{2\pi
c^{2}\hbar^{2}}\mathcal{P}(\wektor{k}_{f})\sum_{\mu}\omega_{\mu}\calkaob{r}\:e^{-i\wektor{k}_{f}\cdot\wektor{r}}
\mathsf{u}_{\mu}(\wektor{r})\nonumber\\
&&\times
\calkaob{r}'\:\mathsf{u}_{\mu}^{\dag}(\wektor{r}')\psi(\wektor{r}'),
\end{eqnarray}
which, by virtue of Eqs. (\ref{I.9}) and (\ref{II.6}), reduces to
\begin{equation}\label{II.8}
\Amplituda=\frac{-E}{2\pi c^{2}\hbar^{2}}
\sum_{s=-}^{+}\Theta_{s}(\wektor{k}_{f})\braket{\wektor{k}_{f}\theta_{s}}{\mathsf{u}}
\mathsf{\Omega}\left(\mathbbm{1}+\mathsf{G}\mathsf{\Omega}\right)^{-1}
\big<\mathsf{u}\big|\phi_{i}\big>
\end{equation}
and, utilizing the fact that for all invertible matrices
$\mathsf{X}$ and $\mathsf{Y}$ the relation
$(\mathsf{X}\mathsf{Y})^{-1}=\mathsf{Y}^{-1}\mathsf{X}^{-1}$ is
satisfied, finally to
\begin{equation}\label{II.9}
\Amplituda=\frac{-E}{2\pi c^{2}\hbar^{2}}
\sum_{s=-}^{+}\Theta_{s}(\wektor{k}_{f})\braket{\wektor{k}_{f}\theta_{s}}{\mathsf{u}}
\left(\mathsf{\Omega}^{-1}+\mathsf{G}\right)^{-1}
\big<\mathsf{u}\big|\phi_{i}\big>.
\end{equation}
Subsequently, using the fact that (\ref{I.11}), we obtain the
bispinor scattering amplitude in the following form
\begin{eqnarray}\label{II.10}
&&\hspace{-1.02cm}\Amplituda=\frac{-E}{2\pi
c^{2}\hbar^{2}}\sum_{s,t=-}^{+}\Theta_{s}(\wektor{k}_{f})\braket{\wektor{k}_{f}\theta_{s}}{\mathsf{u}}\nonumber\\
&&\times \left(\mathsf{\Omega}^{-1}+\mathsf{G}\right)^{-1}
\braket{\mathsf{u}}{\wektor{k}_{i}\theta_{t}}\Theta_{t}^{\dagger}(\wektor{k}_{i})
U_{i}(\wektor{k}_{i}),
\end{eqnarray}
which, after comparison with Eq. (\ref{I.16}), gives the formulae
for the $4\times4$ matrix scattering amplitude:
\begin{eqnarray}\label{II.11}
&&\hspace{-2.02cm}\Amplitudaa=\frac{-E}{2\pi c^{2}\hbar^{2}}
\sum_{s,t=-}^{+}\Theta_{s}(\wektor{k}_{f})\braket{\wektor{k}_{f}\theta_{s}}{\mathsf{u}}\nonumber\\
&&\hspace{-1cm}\times
\left(\mathsf{\Omega}^{-1}+\mathsf{G}\right)^{-1}
\braket{\mathsf{u}}{\wektor{k}_{i}\theta_{t}}\Theta_{t}^{\dagger}(\wektor{k}_{i}),
\end{eqnarray}
and finally, after straightforward movements, for the $2\times 2$
matrix scattering amplitude as
\begin{equation}\label{II.12}
\Amplitudaaa=\frac{-E}{2\pi c^{2}\hbar^{2}}
\sum_{s,t=-}^{+}\theta_{s}
\braket{\wektor{k}_{f}\theta_{s}}{\mathsf{u}}
\left(\mathsf{\Omega}^{-1}+\mathsf{G}\right)^{-1}
\braket{\mathsf{u}}{\wektor{k}_{i}\theta_{t}}\theta_{t}^{\dagger}.
\end{equation}
%

%%%%%%%%%%%%%%%%%%%%%%%%%%%%%%%%%%%%%%%%%%%%%%%%%%%%%%%%%%%%%%
%%%%%%%%%%%%%%                              %%%%%%%%%%%%%%%%%%
%%%%%%%%%%%%%%      Section III             %%%%%%%%%%%%%%%%%%
%%%%%%%%%%%%%%%%%%%%%%%%%%%%%%%%%%%%%%%%%%%%%%%%%%%%%%%%%%%%%%

\section{The eigenchannel method}\label{III}
\setcounter{equation}{0} \noindent\indent Now we are in position
to apply the eigenchannel method proposed recently by Szmytkowski
\cite{Szmytkowski} to scattering of the Dirac particles from
potentials of the form (\ref{II.1}). As we shall see below, such a
class of potentials allows us to formulate this method in a
simplified algebraic form.

We start from the decomposition of the matrix
$\mathsf{\Omega}^{-1}+\mathsf{G}_{\mathrm{D}}$ into its Hermitian
and non-Hermitian parts, i.e.,
\begin{equation}\label{III.1}
\mathsf{\Omega}^{-1}+\mathsf{G}=\mathsf{A}+i\mathsf{B},
\end{equation}
where matrices $\sf{A}$ and $\sf{B}$ are defined through relations
\begin{equation}\label{III.2}
\mathsf{A}=\mathsf{\Omega}^{-1}+\frac{1}{2}
\left(\mathsf{G}+\mathsf{G}^{\dagger}\right),\qquad
%\end{equation}
%
%and
%
%\begin{equation}\label{III.3}
\mathsf{B}=\frac{1}{2i}\left(\mathsf{G}-
\mathsf{G}^{\dagger}\right).
\end{equation}
It is evident from these definitions that both matrices $\sf{A}$
and $\sf{B}$ are Hermitian. Moreover, utilizing the fact that
\begin{equation}\label{III.4}
\wektor{\nabla}\frac{\mathrm{e}^{ik|\wektor{r}-\wektor{r}'|}}{|\wektor{r}-\wektor{r}'|}=
\frac{\wektor{r}-\wektor{r}'}{|\wektor{r}-\wektor{r}'|}\left(\frac{ik\mathrm{e}^{ik|\wektor{r}-\wektor{r}'|}}
{|\wektor{r}-\wektor{r}'|}-\frac{\mathrm{e}^{ik|\wektor{r}-\wektor{r}'|}}{|\wektor{r}-\wektor{r}'|^{2}}\right)
\end{equation}
where $\wektor{\varrho}=\wektor{r}-\wektor{r}'$, the
straightforward calculations lead us to their matrix elements of
the form
\begin{eqnarray}\label{III.6}
&&\mathsf{A}_{\nu\mu}=
\omega^{-1}_{\nu}\delta_{\nu\mu}-\frac{k}{4\pi c^{2}\hbar^{2}}
\calkaob{r}\calkaob{r}'\mathsf{u}_{\nu}^{\dagger}(\wektor{r})\nonumber\\
&&\times\left[ ic\hbar
k\wektor{\alpha}\cdot\frac{\wektor{\varrho}}{|\wektor{\varrho}|}
y_{1}(k|\wektor{\varrho}|)+(\beta
mc^{2}+E)y_{0}(k|\wektor{\varrho}|)
\right]\mathsf{u}_{\mu}(\wektor{r}')\nonumber\\
\end{eqnarray}
and
\begin{eqnarray}\label{III.7}
&&\mathsf{B}_{\nu\mu}=\frac{k}{4\pi c^{2}\hbar^{2}}
\calkaob{r}\calkaob{r}'\mathsf{u}_{\nu}^{\dagger}(\wektor{r})\nonumber\\
&&\times\left[ic\hbar
k\wektor{\alpha}\cdot\frac{\wektor{\varrho}}{|\wektor{\varrho}|}
j_{1}(k|\wektor{\varrho}|)+(\beta
mc^{2}+E)j_{0}(k|\wektor{\varrho}|)
\right]\mathsf{u}_{\mu}(\wektor{r}'),\nonumber\\
\end{eqnarray}
where $j_{0}(z)$, $j_{1}(z)$, $y_{0}(z)$ and $y_{1}(z)$ are,
respectively, the Bessel and Neumann spherical functions
\cite{AbrStegun}. Recall that in general
$j_{0}(z)=(\sin z)/z$,
$y_{0}(z)=(\cos z)/z$
and
\begin{equation}
j_{1}(z)=-\frac{\cos z}{z}+\frac{\sin z}{z^{2}}, \quad
y_{1}(z)=-\frac{\sin z}{z}-\frac{\cos z}{z^{2}}.
\end{equation}
The main idea of the present paper, adopted from
\cite{Szmytkowski}, is to construct the following weighted
spectral problem:
\begin{equation}\label{III.8}
\mathsf{A}X_{\gamma}(E)=\lambda_{\gamma}(E)\mathsf{B}
X_{\gamma}(E),
\end{equation}
where $X_{\gamma}(E)$ and $\lambda_{\gamma}(E)$ is, respectively,
an eigenvector and an eigenvalue. Thereafter the eigenvectors
$\{X_{\gamma}(E)\}$ will be called {\it eigenchannel vectors}.
They are directly related to eigenchannels defined in
\cite{Szmytkowski} as state vectors. In fact, they constitute a
projection of eigenchannels onto subspace spanned by
$\mac{u}_{\mu}(\wektor{r})$.

Using the fact that matrices $\mathsf{A}$ and $\mathsf{B}$ are
Hermitian and, as it is proven in Appendix \ref{AppA}, the matrix
$\mathsf{B}$ is positive semi-definite, one finds that the
eigenvalues $\{\lambda_{\gamma}(E)\}$ are real, i.e.,
$\lambda_{\gamma}^{*}(E)=\lambda_{\gamma}(E)$. Moreover, the
eigenchannels associated with different eigenvalues obey the
orthogonality relation
\begin{equation}\label{III.9}
X_{\gamma'}^{\dagger}(E)\mathsf{B}X_{\gamma}(E)=0 \qquad
(\lambda_{\gamma'}(E)\neq\lambda_{\gamma}(E)).
\end{equation}
In case of degeneration of some eigenvalues one may always choose
the corresponding eigenvectors to be orthogonal according to the
above relation. Then, imposing the normalization
$X_{\gamma}^{\dagger}(E)\mathsf{B}X_{\gamma}(E)=1$, one obtains
the following orthonormality relation
\begin{equation}\label{III.10}
X_{\gamma'}^{\dagger}(E)\mathsf{B}X_{\gamma}(E)=\delta_{\gamma'\gamma}.
\end{equation}
From Eqs. (\ref{III.8}) and (\ref{III.10}) one infers that the
eigenvalues $\{\lambda_{\gamma}(E)\}$ may be related to the matrix
$\mathsf{A}$ as follows
\begin{equation}\label{III.11}
\lambda_{\gamma}(E)=X_{\gamma}^{\dagger}(E)\mathsf{A}X_{\gamma}(E).
\end{equation}
Similar reasoning may be carried out employing the matrices
$\mathsf{A}$ and $\mathsf{\Omega}^{-1}+\mathsf{G}$. Indeed, after
algebraic manipulations we arrive at
\begin{eqnarray}\label{III.12}
&&X_{\gamma'}^{\dagger}(E)\mathsf{A}X_{\gamma}(E)=
\lambda_{\gamma}(E)\delta_{\gamma'\gamma},\nonumber\\
&&\hspace{-1cm}X_{\gamma'}^{\dagger}(E)(\mathsf{\Omega}^{-1}+\mathsf{G})X_{\gamma}(E)=
[i+\lambda_{\gamma}(E)]\delta_{\gamma'\gamma},
\end{eqnarray}
and
%
%\begin{equation}\label{III.13}
$\lambda_{\gamma}(E)=X_{\gamma}^{\dagger}(E)(\mathsf{\Omega}^{-1}+\mathsf{G})X_{\gamma}(E)-i$.
%\end{equation}
%
Since the eigenchannels $\{X_{\gamma}(E)\}$ are the solutions of
the Hermitian eigenvalue problem, they may satisfy the following
closure relations
\begin{eqnarray}\label{III.14}
&\displaystyle \sum_{\gamma}X_{\gamma}(E)X_{\gamma}^{\dagger}(E)\mathsf{B}=\jed,&\nonumber\\
&\displaystyle \sum_{\gamma}\lambda_{\gamma}^{-1}(E)
X_{\gamma}(E)X_{\gamma}^{\dagger}(E)\mathsf{A}=\jed,&
\end{eqnarray}
and
\begin{equation}\label{III.15}
\sum_{\gamma}
\frac{1}{i+\lambda_{\gamma}(E)}X_{\gamma}(E)X_{\gamma}^{\dagger}(E)
(\mathsf{\Omega}^{-1}+\mathsf{G})=\jed,
\end{equation}
where $\jed$ is an identity matrix, which dimension depends on the
dimension of the matrix $\mathsf{G}$. For purposes of further
analyzes the above closure relations are assumed to hold. Below, we
employ the above reasoning to the derivation of the scattering
amplitudes. From Eq. (\ref{III.15}) one deduces that
\begin{equation}\label{III.16}
(\mathsf{\Omega}^{-1}+\mathsf{G})^{-1}=\sum_{\gamma}
\frac{1}{i+\lambda_{\gamma}(E)}
X_{\gamma}(E)X_{\gamma}^{\dagger}(E).
\end{equation}
After substitution of Eq. (\ref{III.16}) to Eq. (\ref{II.11}) and
rearranging terms, we have
\begin{eqnarray}\label{III.17}
&&\hspace{-0.5cm}\Amplitudaa=\frac{-E}{2\pi c^{2}\hbar^{2}}
\sum_{\gamma}\frac{1}{i+\lambda_{\gamma}(E)}\sum_{s=-}^{+}\Theta_{s}(\wektor{k}_{f})
\braket{\wektor{k}_{f}\theta_{s}}{\mathsf{u}}X_{\gamma}(E)\nonumber\\
&&\hspace{0.5cm}\times \sum_{t=-}^{+}X_{\gamma}^{\dagger}(E)
\braket{\mathsf{u}}{\wektor{k}_{i}\theta_{t}}\Theta_{t}^{\dagger}(\wektor{k}_{i}).
\end{eqnarray}
Let us define the following angular functions
\begin{equation}\label{III.18}
\mathcal{Y}_{\gamma}(\wektor{k})=\sqrt{\frac{Ek}{8\pi^{2}c^{2}\hbar^{2}}}
\sum_{s=-}^{+}\Theta_{s}(\wektor{k})\braket{\wektor{k}\theta_{s}}{\mathsf{u}}
X_{\gamma}(E),
\end{equation}
hereafter termed the {\it eigenchannel bispinor harmonics}. The
functions $\{\mathcal{Y}_{\gamma}(\wektor{k})\}$ are orthonormal
on the unit sphere (for proof, see Appendix \ref{AppB}), i.e.,
\begin{equation}\label{III.19}
\calkapow{k}\,\mathcal{Y}_{\gamma'}^{\dagger}(\wektor{k})\mathcal{Y}_{\gamma}(\wektor{k})
=\delta_{\gamma'\gamma}.
\end{equation}
Application of Eq. (\ref{III.18}) to Eq. (\ref{III.17}) yields
\begin{equation}\label{III.20}
\Amplitudaa=\frac{4\pi}{k}\sum_{\gamma}e^{i\delta_{\gamma}(E)}\sin\delta_{\gamma}(E)
\mathcal{Y}_{\gamma}(\wektor{k}_{f})\mathcal{Y}_{\gamma}^{\dagger}(\wektor{k}_{i}),
\end{equation}
where $\{\delta_{\gamma}(E)\}$ are called {\it eigenphase-shifts}
and are related to $\{\lambda_{\gamma}(E)\}$ according to
\begin{equation}\label{III.21}
\lambda_{\gamma}(E)=-\cot\delta_{\gamma}(E).
\end{equation}
Similar considerations may be carried out for the $2\times 2$
matrix scattering amplitude $\Amplitudaaa$. Indeed, in virtue of
Eq. (\ref{I.17}) we may rewrite Eq. (\ref{II.12}) in the form
\begin{equation}\label{III.22}
\Amplitudaaa=\frac{4\pi}{k}\sum_{\gamma}e^{i\delta_{\gamma}(E)}
\sin\delta_{\gamma}(E)\Upsilon_{\gamma}(\wektor{k}_{f})\Upsilon_{\gamma}^{\dagger}(\wektor{k}_{i}),
\end{equation}
where the angular functions $\{\Upsilon_{\gamma}(\wektor{k})\}$,
hereafter called {\it eigenchannel spinor harmonics}, are defined as
follows
\begin{equation}\label{III.23}
\Upsilon_{\gamma}(\wektor{k})=\sqrt{\frac{Ek}{8\pi^{2}c^{2}\hbar^{2}}}
\sum_{s=-}^{+}\theta_{s}\braket{\wektor{k}\theta_{s}}{\mathsf{u}}
X_{\gamma}(E).
\end{equation}
Moreover, they are orthogonal on the unit sphere (for proof, see
Appendix \ref{AppB})
\begin{equation}\label{III.24}
\calkapow{k}\,\Upsilon_{\gamma'}^{\dagger}(\wektor{k})\Upsilon_{\gamma}(\wektor{k})=\delta_{\gamma'\gamma},
\end{equation}
and, as one can verify, are related to the eigenchannel bispinor
harmonics $\{\mathcal{Y}_{\gamma}(\wektor{k})\}$ via the relation
\begin{equation}\label{III.25}
\mathcal{Y}_{\gamma}(\wektor{k})=\frac{1}{\sqrt{1+\varepsilon^{2}}}
\left(
\begin{array}{c}
\Upsilon_{\gamma}(\wektor{k})\\
\varepsilon\wektor{\sigma}\cdot\wersor{k}\Upsilon_{\gamma}(\wektor{k})
\end{array}
\right).
\end{equation}
Now we are in position to compute scattering cross-sections.
Substitution of Eq. (\ref{III.22}) to Eq. (\ref{I.18}) and
integration over all directions of scattering
$\hat{\wektor{k}}_{f}$, by virtue of relation (\ref{III.24}),
yields
\begin{equation}\label{III.26}
\sigma(\wektor{k}_{i},\wektor{\nu}_{i})=\frac{16\pi^{2}}{k^{2}}
\sum_{\gamma}\sin^{2}\delta_{\gamma}(E)\left|\chi_{i}^{\dagger}\Upsilon_{\gamma}(\wektor{k}_{i})\right|^{2}.
\end{equation}
To compute the total cross-section averaged over all arrangements
of spin of the incident particle, we have to notice that the
projector onto the pure state $\chi_{i}$ may be written as
$\chi_{i}\chi_{i}^{\dagger}=(1/2)[\jed_{2}+\wektor{\nu}_{i}\cdot\wektor{\sigma}]$
with $|\wektor{\nu}_{i}|=1$.
Therefore, substituting of the above to Eq. (\ref{III.26}) and
averaging over all directions of $\boldsymbol{\nu}_{i}$, we arrive
at
\begin{equation}\label{III.28}
\sigma(\wektor{k}_{i})=
\frac{8\pi^{2}}{k^{2}}\sum_{\gamma}\sin^{2}\delta_{\gamma}(E)\Upsilon_{\gamma}^{\dagger}(\wektor{k}_{i})
\Upsilon_{\gamma}(\wektor{k}_{i}).
\end{equation}
Finally, averaging the above scattering cross-section over all
directions of incidence $\hat{\wektor{k}}_{i}$, again by virtue of
Eq. (\ref{III.24}), we get the total cross-section in the form
\begin{equation}\label{III.29}
\sigma_{t}(E)=\frac{2\pi}{k^{2}}\sum_{\gamma}\sin^{2}\delta_{\gamma}(E).
\end{equation}
It should be emphasized that all the above considerations
respecting scattering cross-sections may be repeated using the
eigenchannel bispinor harmonics
$\{\mathcal{Y}_{\gamma}(\wektor{k})\}$ instead of the eigenchannel
spinor harmonics $\{\Upsilon_{\gamma}(\wektor{k})\}$. The
significant difference is that then the integrals over
$\hat{\wektor{k}}_{f}$ and $\hat{\wektor{k}}_{i}$ need to be
calculated using relation (\ref{III.19}) instead of
(\ref{III.24}).

%%%%%%%%%%%%%%%%%%%%%%%%%%%%%%%%%%%%%%%%%%%%%%%%%%%%%%%%%%%%%%%%%%%%%
%%%%%%%%%%%%%%%%%%%                             %%%%%%%%%%%%%%%%%%%%%
%%%%%%%%%%%%%%%%%%%     Section IV              %%%%%%%%%%%%%%%%%%%%%
%%%%%%%%%%%%%%%%%%%%%%%%%%%%%%%%%%%%%%%%%%%%%%%%%%%%%%%%%%%%%%%%%%%%%

\section{Example}
\setcounter{equation}{0}
We conclude our considerations providing here an illustrative
example concerning the scattering from a spherical shell of radius
$R$, centered at the origin of the coordinate system. Due to the
assumption of non-locality of potentials under consideration, we
shall simulate this process by using a potential of the form
\begin{equation}\label{Ex1}
\mac{V}(\wektor{r},\wektor{r}')=\omega
v(\wektor{r})v(\wektor{r}')\jed_{4},\qquad
v(\wektor{r})=\frac{1}{\sqrt{4\pi}}\frac{\delta(r-R)}{R^{2}},
\end{equation}
where $\omega\neq 0$. Notice that the potential defined above is
the special case of that proposed recently by de Prunel\'e
\cite{Prunele} (see also \cite{Prunele1}). Scattering of the Dirac
particles from delta-like potentials was also studied e.g. in
Refs. \cite{Dombey,Loewe}. However, in these papers the authors
considered only local potentials and not non-local ones.

At the very beginning, we need to bring the potential (\ref{Ex1}) to
the previously postulated form (\ref{II.1}). To this aim, let
$\mac{e}_{1}$ and $\mac{e}_{2}$ constitute a standard basis in
$\mathbb{C}^{2}$, i.e., $\mac{e}_{1}=(1\;0)^{T}$ and
$\mac{e}_{2}=(0\;1)^{T}$. Moreover, let
$\mac{e}_{ij}=\mac{e}_{i}\otimes\mac{e}_{j}$ and then by virtue of
the fact that
$\jed_{4}=\sum_{i,j=1}^{2}\mac{e}_{ij}\mac{e}_{ij}^{\dagger}$, we
may rewrite (\ref{Ex1}) as
\begin{equation}\label{Ex2}
\mac{V}(\wektor{r},\wektor{r}')=\omega\sum_{i,j=1}^{2}\mac{u}_{ij}(\wektor{r})\mac{u}_{ij}^{\dagger}(\wektor{r}),\qquad
\mac{u}_{ij}(\wektor{r})=v(\wektor{r})\mac{e}_{ij}.
\end{equation}
Now, we are in position to compute the matrix $\mac{G}$. Using
Eqs. (\ref{I.6}) and (\ref{Ex1}), after straightforward
integrations we have
\begin{eqnarray}\label{Ex3}
\mac{G}=ikj_{0}(kR)h_{0}^{(+)}(kR) \left(
\begin{array}{cc}
\eta_{+}\jed_{2} & 0\\
0 & \eta_{-}\jed_{2}
\end{array}
\right),
\end{eqnarray}
where $\eta_{\pm}=(E\pm mc^{2})/c^{2}\hbar^{2}$ and
$h_{0}^{(+)}(z)=j_{0}(z)+iy_{0}(z)$ is the spherical Hankel
function of the first kind. Hence, by the definitions given in Eq.
(\ref{III.2}), we find that the explicit forms of matrices
$\mac{A}$ and $\mac{B}$ are
\begin{equation}\label{Ex4}
\mac{A}=\left(
\begin{array}{cc}
[\omega^{-1}-kj_{0}(kR)y_{0}(kR)\eta_{+}]\jed_{2} &\hspace{-1.7cm} 0\\
0 & \hspace{-1.7cm}
[\omega^{-1}-kj_{0}(kR)y_{0}(kR)\eta_{-}]\jed_{2}
\end{array}
\right)
\end{equation}
and
\begin{equation}\label{Ex5}
\mac{B}=kj_{0}^{2}(kR) \left(
\begin{array}{cc}
\eta_{+}\jed_{2} & 0\\
0 & \eta_{-}\jed_{2}
\end{array}
\right).
\end{equation}
According to the method formulated in Sec. \ref{III}, we may
construct the following spectral problem
\begin{equation}\label{Ex6}
\mac{A}X_{\gamma}(E)=\lambda_{\gamma}(E)\mac{B}X_{\gamma}(E)\qquad
(\gamma=1,2,3,4),
\end{equation}
which, as one can easily verify, has two different eigenvalues
\begin{equation}\label{Ex7}
\lambda_{\pm}(E)=\frac{\omega^{-1}-kj_{0}(kR)y_{0}(kR)\eta_{\pm}}{kj_{0}^{2}(kR)\eta_{\pm}}
\end{equation}
and respective eigenvectors
\begin{eqnarray}\label{Ex8}
&\displaystyle X_{+}^{(1(2))}(E)=\frac{1}{\sqrt{k\eta_{+}}j_{0}(kR)}\,\mac{e}_{1}\otimes\mac{e}_{1(2)},&\nonumber\\
&\displaystyle X_{-}^{(1(2))}(E)=\frac{1}{\sqrt{k\eta_{-}}j_{0}(kR)}
\,\mac{e}_{2}\otimes\mac{e}_{1(2)}.&
\end{eqnarray}
Then, using Eq. (\ref{III.18}) and by virtue of the fact that
\begin{eqnarray}\label{Ex9}
&&\hspace{-1.5cm}\braket{\wektor{k}\chi}{\mac{\bf{u}}}=\sqrt{\frac{4\pi}{1+\varepsilon^{2}}}j_{0}(kR) \nonumber\\
&&\hspace{-1cm}\times \left( \chi^{\dagger}\mac{e}_{1}\;\;
\chi^{\dagger}\mac{e}_{2}\;\;\varepsilon\chi^{\dagger}\wektor{\sigma}\cdot\wersor{k}\,
\mac{e}_{1}\;\;
\varepsilon\chi^{\dagger}\wektor{\sigma}\cdot\wersor{k}\,
\mac{e}_{2}\right),
\end{eqnarray}
we arrive at the four eigenchannel bispinor harmonics
$\{\mathcal{Y}_{\gamma}(\wektor{k})\}$ in the form
\begin{equation}\label{Ex10}
\mathcal{Y}_{+}^{(1(2))}(\wektor{k})=
\frac{1}{\sqrt{4\pi(1+\varepsilon^{2})}}\, \left(
\begin{array}{c}
\mac{e}_{1(2)}\\
\varepsilon\wektor{\sigma}\cdot\wersor{k}\,\mac{e}_{1(2)}
\end{array}
\right)
\end{equation}
and
\begin{equation}
\mathcal{Y}_{-}^{(1(2))}(\wektor{k})=\frac{1}{\sqrt{4\pi(1+\varepsilon^{2})}}\,
\left(
\begin{array}{c}
\varepsilon\wektor{\sigma}\cdot\wersor{k}\,\mac{e}_{1(2)}\\
\mac{e}_{1(2)}
\end{array}
\right).
\end{equation}
Then, by virtue of Eq. (\ref{III.23}), one obtains the eigenchannel
spinor harmonics $\{\Upsilon_{\gamma}(\wektor{k})\}$ in the form
\begin{equation}\label{Ex11}
\Upsilon_{+}^{(1(2))}(\wektor{k})=
\frac{1}{\sqrt{4\pi}}\,\mac{e}_{1(2)},\quad
\Upsilon_{-}^{(1(2))}(\wektor{k})=\frac{1}{\sqrt{4\pi}}\,\wektor{\sigma}\cdot\wersor{k}\,\mac{e}_{1(2)}.
\end{equation}
The latter may be equivalently obtained combining Eqs.
(\ref{III.25}) and (\ref{Ex11}). Moreover, as one may easily
verify, functions given by Eqs. (\ref{Ex10}) and (\ref{Ex11}) are
orthonormal, respectively, in the sense (\ref{III.19}) and
(\ref{III.24}).

Before we find an expression for total cross section, we compute
the scattering amplitude. Since, as shown in Sec. \ref{SecII}, the
bispinor and both matrix scattering amplitudes are mutually
related, we restrict our considerations to the $2\times 2$
scattering amplitude. Thus, combining Eqs. (\ref{III.22}),
(\ref{Ex5}), and (\ref{Ex11}) we obtain

\begin{eqnarray}\label{Ex12}
&&\hspace{-0.9cm}\Amplitudaaa=-j_{0}^{2}(kR)\left[
\frac{\jed_{2}}{ik
j_{0}(kR)h_{0}^{(+)}(kR)+(\omega\eta_{+})^{-1}}\right.\nonumber\\
&&\left.+\frac{(\wektor{\sigma}\cdot\wersor{k}_{f})
(\wektor{\sigma}\cdot\wersor{k}_{i})}
{ikj_{0}(kR)h_{0}^{(+)}(kR)+(\omega\eta_{-})^{-1}} \right].
\end{eqnarray}
Finally, substitution of Eqs. (\ref{Ex8}) and (\ref{Ex11}) to Eq.
(\ref{III.26}) with the aid of Eq. (\ref{III.21}) yields
\begin{eqnarray}\label{Ex13}
&&\hspace{-1cm}\sigma(\wektor{k}_{i},\wektor{\nu}_{i})=\frac{4\pi}{k^{2}}
j_{0}^{4}(kR)\nonumber\\
&&\hspace{-0.5cm}\times\left\{
\frac{1}{[(k\omega \eta_{+})^{-1}-j_{0}(kR)y_{0}(kR)]^{2}+j_{0}^{4}(kR)}\right.\nonumber\\
&&\hspace{-0.5cm}+\left.\frac{1}{[(k\omega\eta_{-})^{-1}-j_{0}(kR)y_{0}(kR)]^{2}
+j_{0}^{4}(kR)}\right\}.
\end{eqnarray}
Here it is evident that
$\sigma(\wektor{k}_{i},\wektor{\nu}_{i})=\sigma(\wektor{k}_{i})=\sigma_{t}(E)$.

In order to illustrate the obtained results, the eigenphaseshifts
for two different values of $\omega$, derived from Eqs.
(\ref{III.21}) and (\ref{Ex7}), are plotted in Fig. 1 and 2.
Figures 3 and 4 present partial $\sigma_{\pm}(E)$ as well as total
$\sigma_{t}(E)$ cross sections.

It seems interesting to investigate the behavior of both eigenvalues
$\lambda_{\pm}(E)$ in the non-relativistic limit, i.e., when
$c\to\infty$. From (\ref{DiracWaveNumber}) one concludes that
\begin{equation}
\eta_{+}\stackrel{c\to\infty}{\LRA}\frac{2m}{\hbar^{2}},\qquad
\eta_{-}\stackrel{c\to\infty}{\LRA}0
\end{equation}
and therefore
\begin{equation}
\lambda_{+}(E)\stackrel{c\to\infty}{\LRA}\frac{(\hbar^{2}/2m\omega)-kj_{0}(kR)y_{0}(kR)}{kj_{0}^{2}(kR)}
\end{equation}
and
\begin{equation}
\lambda_{-}(E)\stackrel{c\to\infty}{\LRA}\mathrm{sgn}(\omega)\infty.
\end{equation}
This means that $\delta_{-}(E)\to n\pi$ $(n\in \mathbb{Z})$ in the
limit of $c\to\infty$. Therefore the cross section $\sigma_{-}(E)$
vanishes in the non-relativistic limit  and in this sense it has a
purely relativistic character leading to the fact that the
resonance appearing in Fig. 4 at about $1.25 mc^{2}$ is purely
relativistic effect.

One sees that in the non-relativistic limit the cross section
(\ref{Ex13}) reduces to
\begin{eqnarray}
&&\hspace{-1cm}\sigma_{t}(E)\stackrel{c\to\infty}{\LRA}\frac{4\pi}{k^{2}}
j_{0}^{4}(kR)\nonumber\\
&&\hspace{-0.8cm}\times\left\{ \frac{1}{[(\hbar^{2}/2m
k\omega)-j_{0}(kR)y_{0}(kR)]^{2}+j_{0}^{4}(kR)}\right\}.
\end{eqnarray}
The above cross section may also be obtained using non-relativistic
formulation of the present method given in Ref. \cite{moja}.

\begin{figure}[ht]
\includegraphics[width=6cm]{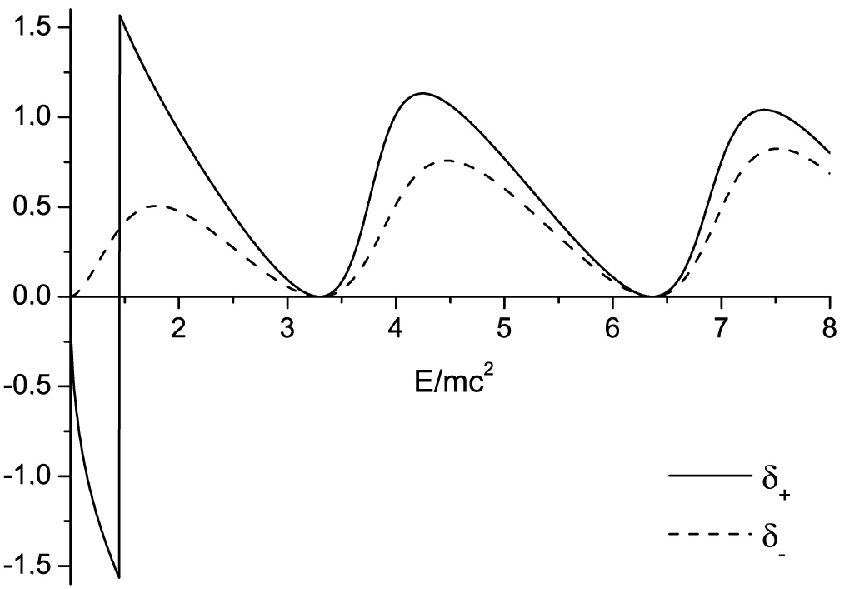}
\caption{Behavior of eigenphaseshifts $\delta_{+}(E)$ (solid
curve) and $\delta_{-}(E)$ (dashed curve) as functions of energy
$E$ (in units of $mc^{2}$) for $\omega=-\hbar^{3}/m^{2}c$ and
$R=\hbar/mc$. The eigenphaseshift $\delta_{+}(E)$ has been
constrained to the range $[-\pi/2,\pi/2]$. }
\end{figure}
\begin{figure}[ht]
\includegraphics[width=6cm]{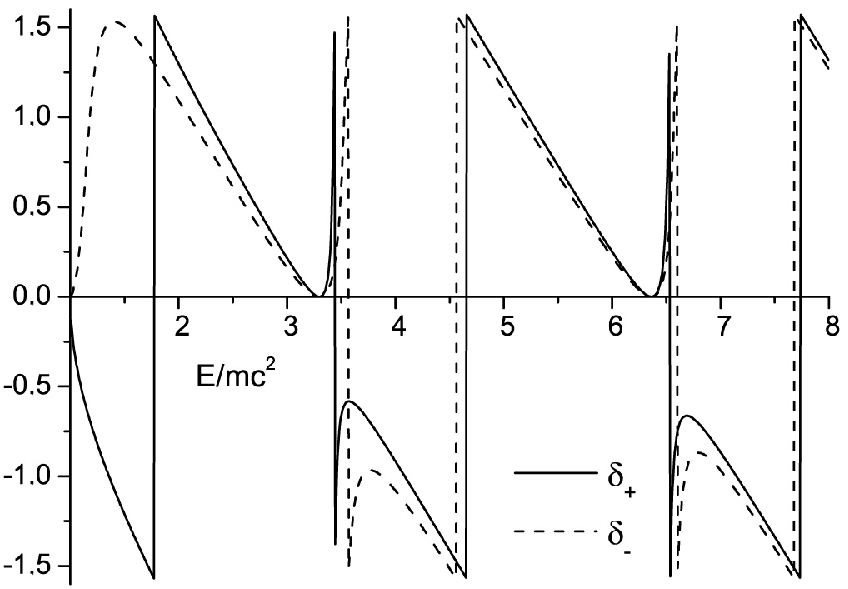}
\caption{Behavior of eigenphaseshifts $\delta_{+}(E)$ (solid curve)
and $\delta_{-}(E)$ (dashed curve) as functions of energy $E$ (in
units of $mc^{2}$) for $\omega=-5\hbar^{3}/m^{2}c$ and $R=\hbar/mc$.
Both eigenphaseshifts have been constrained to the range
$[-\pi/2,\pi/2]$.}
\end{figure}
\begin{figure}[ht]
\includegraphics[width=6cm]{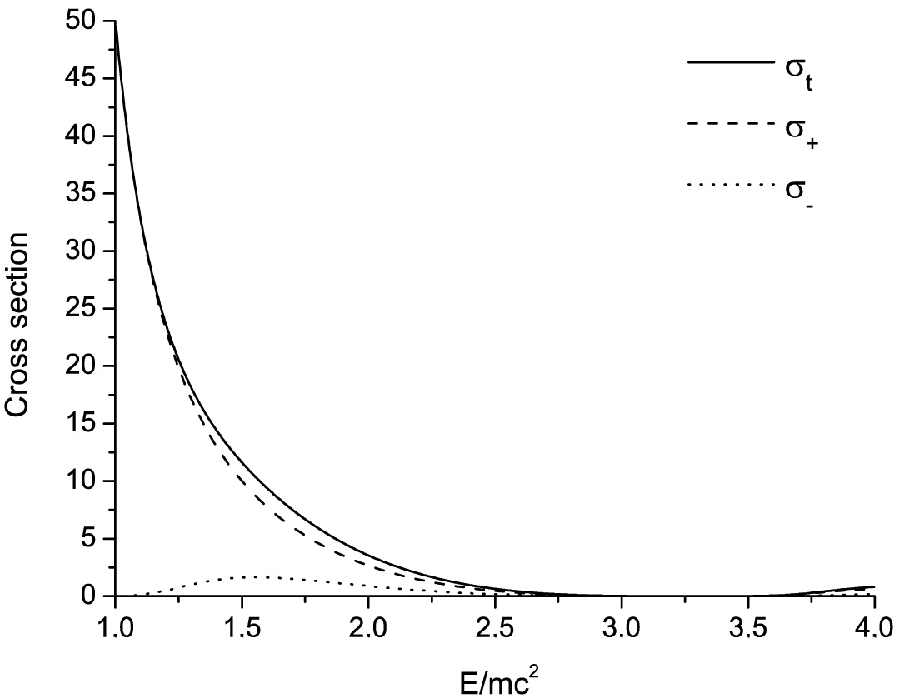}
\caption{Partial $\sigma_{+}(E)$ (dashed curve), $\sigma_{-}(E)$
(dotted curve), and total $\sigma_{t}(E)$ (solid curve) cross
sections (all in units of $R^{2}$) as functions of energy $E$ (in
units of $mc^{2}$) for $\omega=-\hbar^{3}/m^{2}c$ and
$R=\hbar/mc$.}
\end{figure}

\begin{figure}[ht]
\includegraphics[width=6cm]{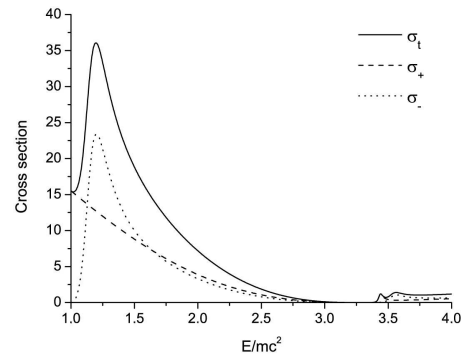}
\caption{Partial $\sigma_{+}(E)$ (dashed curve), $\sigma_{-}(E)$
(dotted curve), and total $\sigma_{t}(E)$ (solid curve) cross
sections (all in units of $R^{2}$) as functions of energy $E$ (in
units of $mc^{2}$) for $\omega=-5\hbar^{3}/m^{2}c$ and
$R=\hbar/mc$. }
\end{figure}

\section{Conclusions}
In this work, an application of the recently proposed eigenchannel
method \cite{Szmytkowski} to the scattering of Dirac particles from
non-local separable potentials has been presented. Application of
such a particular case of the non-local potentials reduces naturally
the general weighted eigenvalue problem stated in Ref.
\cite{Szmytkowski} to its matrix counterpart given by Eq.
(\ref{III.8}) leading to the definition of eigenchannel vectors.
Using the notion of the eigenchannel vectors the definitions of
eigenchannel spinor as well as bispinor harmonics have been given.
The latter provide us with the formulas for scattering amplitudes
similar to that well-known for central potentials generalizing
them at the same time to the case of non-local separable
potentials.

The general considerations have been extended with an illustrative
example in which the Dirac particles are scattered from non-local,
delta-like potential. In this particular case, the general
eigenvalue problem (\ref{III.8}) become just a $4\times 4$ matrix
equation and therefore is easily solvable (notice that in the case
of non-relativistic scattering it would be just a one-dimensional
problem). The eigenvalues of this problem are two-fold degenerated
and therefore give two different eigenphase-shifts from which one
has a purely relativistic character in the sense that it tends to
$n\pi$ $(n\in \mathbb{Z})$ whenever $c\to\infty$ giving no
contribution to total cross sections in non-relativistic limit.
One sees also that even such a simple example of non-local
potentials may lead to some resonances (see Fig. 4).

The next step in our considerations will be to investigate the
applicability of the new formulation of the eigenchannel method in
the case of inelastic scattering from separable potentials.
Moreover it seems also interesting to investigate the
applicability of the method to the other, more complicated
examples of separable potentials.

\section*{Acknowledgments}

I am grateful to R.~Szmytkowski for very useful discussions,
suggestions and commenting on the manuscript. Discussions with
M.~Czachor are also acknowledged.

\appendix
\renewcommand{\theequation}{\Alph{section}.\arabic{equation}}

\section{Positive semidefiniteness of the matrix $\mac{B}$}
\label{AppA} \setcounter{equation}{0}

The proof follows the suggestions of Szmytkowski \cite{Szmytkowski4}. Positive
semidefiniteness of the matrix $\mac{B}$ means that the inequality
\begin{equation}\label{AppA1}
X_{\gamma}^{\dagger}(E)\mac{B}X_{\gamma}(E)\ge 0
\end{equation}
is satisfied. To prove the above statement let us notice that
\begin{eqnarray}\label{AppA2}
&&\hspace{-1cm}\displaystyle\calkapow{k}\:e^{i\wektor{k}\cdot\wektor{\varrho}}
(c\hbar\wektor{\alpha}\cdot\wektor{k}+\beta mc^{2}+E\mathbbm{1}_{4})= 4\pi
\nonumber\\
&&\hspace{-0.7cm}\times\left[ ic\hbar kj_{1}(k|\wektor{\varrho}|)
\wektor{\alpha}\cdot\frac{\wektor{\varrho}}{|\wektor{\varrho}|}+(\beta
mc^{2}+E\mathbbm{1}_{4}) j_{0}(k|\wektor{\varrho}|) \right],
\end{eqnarray}
where $\wektor{\varrho}=\wektor{r}-\wektor{r}'$. Then using Eq.
(\ref{I.8}), we may rewrite Eq. (\ref{III.7}) in the form
\begin{eqnarray}\label{AppA3}
&&\hspace{-0.3cm}\mac{B}_{\nu\mu}=\frac{Ek}{8\pi^{2}c^{2}\hbar^{2}}\calkapow{k}
\calkaob{r}\,e^{i\wektor{k}\cdot\wektor{r}}\mac{u}_{\nu}^{\dagger}(\wektor{r})
\mathcal{P}(\wektor{k})\nonumber\\
&&\hspace{1cm}\times\calkaob{r}'\,e^{-i\wektor{k}\cdot\wektor{r}'}
\mac{u}_{\mu}(\wektor{r}'),
\end{eqnarray}
which after application to Eq. (\ref{AppA1}) yields
\begin{eqnarray}\label{AppA4}
&&X_{\gamma}^{\dagger}(E)\mac{B}X_{\gamma}(E)=
\frac{mk}{8\pi^{2}\hbar^{2}}\nonumber\\
&&\times\calkapow{k}
\left|\left|\sum_{\nu}X_{\gamma\nu}^{*}(E)\calkaob{r}\,e^{i\wektor{k}\cdot\wektor{r}}
\mac{u}_{\nu}^{\dagger}(\wektor{r})\mathcal{P}(\wektor{k})\right|\right|^{2}\ge
0,\nonumber\\
\end{eqnarray}
finishing obviously the proof. Here $X_{\gamma\nu}(E)$ denotes the
$\nu$th element of the eigenchannel vector $X_{\gamma}(E)$ and $||\Omega||=\sqrt{\Omega^{\dag}\Omega}$.

\section{Orhtonormality of the angular functions
$\mathcal{Y}_{\gamma}(\wektor{k})$ and
$\Upsilon_{\gamma}(\wektor{k})$ } \label{AppB}

We begin with proof for the functions
$\mathcal{Y}_{\gamma}(\wektor{k})$. Application of Eq.
(\ref{III.18}) to Eq. (\ref{III.19}) with the aid of Eq.
(\ref{I.9}) and the fact that $\mathcal{P}(\wektor{k})$ is a
projector, we can deduce that
\begin{eqnarray}\label{AppB1}
&&\hspace{-0.5cm}\calkapow{k}\,\mathcal{Y}_{\gamma'}^{\dagger}(\wektor{k})\mathcal{Y}_{\gamma}(\wektor{k})
=\frac{Ek}{8\pi^{2}c^{2}\hbar^{2}}\sum_{\nu\mu}X_{\gamma'\nu}^{*}(E)\nonumber\\
&&\times\left[\calkaob{r}\calkaob{r}'\mathsf{u}_{\nu}^{\dagger}(\wektor{r})\right.\nonumber\\
&&\times\left.
\calkapow{k}\,e^{i\wektor{k}\cdot(\wektor{r}-\wektor{r}')}\mathcal{P}(\wektor{k})
\mathsf{u}_{\mu}(\wektor{r}')\right]X_{\gamma\mu}(E),
\end{eqnarray}
Comparison with Eq. (\ref{AppA3}) shows that the square brackets
in the above contain the expression proportional to certain
element of the matrix $\mac{B}$. Therefore, we may
rewrite Eq. (\ref{AppB1}) as
\begin{equation}\label{AppB2}
\calkapow{k}\,\mathcal{Y}_{\gamma'}^{\dagger}(\wektor{k})\mathcal{Y}_{\gamma}(\wektor{k})=
X_{\gamma'}^{\dagger}(E)\mathsf{B}X_{\gamma}(E).
\end{equation}
Finally, substitution of Eq. (\ref{III.10}) to Eq. (\ref{AppB2})
leads us directly to Eq. (\ref{III.19}), finishing the proof. To
prove the orthonormality relation for the functions
$\{\Upsilon_{\gamma}(\wektor{k})\}$, it suffices to combine Eq.
(\ref{III.25}) with Eq. (\ref{AppB2}).

\end{document}